\documentclass[aps,prd,reprint,groupedaddress,preprintnumbers]{revtex4-1}
\pdfoutput=1
\usepackage{hyperref}
\usepackage{color}

\newcommand{\gev}{\ensuremath{\hbox{ GeV}}}

\newcommand{\jpsi}{\ensuremath{J\!/\!\psi}}

\newcommand{\T}{\rule{0pt}{2.6ex}}
\newcommand{\B}{\rule[-1.2ex]{0pt}{0pt}} 
\newcommand{\alphas}{\ensuremath{\alpha_{\mathrm{s}}}}

\begin{document}

\preprint{FERMILAB--PUB--19/176--T}

\title{Quarkonium wave functions at the origin: an update}


\author{Estia J. Eichten}
\email[]{eichten@fnal.gov}
\affiliation{Theoretical Physics Department, Fermi National Accelerator Laboratory, P.O. Box 500, Batavia, Illinois 60510 USA}
\author{Chris Quigg}
\email[]{quigg@fnal.gov}
\affiliation{Theoretical Physics Department, Fermi National Accelerator Laboratory, P.O. Box 500, Batavia, Illinois 60510 USA}


\date{\today}

\begin{abstract}
Using a newly developed interquark potential, we tabulate values of the radial Schr\"{o}dinger wave function or its first nonvanishing derivative at zero quark--antiquark separation, for $c\bar{c}$, $c\bar{b}$, and $b\bar{b}$ levels that lie below, or just above, the flavor threshold. These quantities are required inputs for evaluating quarkonium production cross sections.
\end{abstract}


\maketitle

Fragmentation of partons into quarkonium states has long been recognized as an important component of quarkonium production in high-energy collisions~\cite{Braaten:1996pv, *[{}][{; \S4 of }]Kramer:2001hh, *Brambilla:2010cs}. It is an essential element of the event generators now in common use to simulate $\jpsi, \Upsilon$, and $B_c$ production~\cite{[{}][{, \url{http://helac-phegas.web.cern.ch/helac-phegas/helac-onia.html}. For integration into PYTHIA, see  \url{http://home.thep.lu.se/~torbjorn/pythia82html/HelacOniaProcesses.html}; }]Shao:2015vga, *Chang:2015qea}.

The calculation of the production rate by fragmentation factorizes into a parton-level piece that can be evaluated using perturbative techniques for Quantum Chromodynamics times a hadronic piece expressed in terms of the quarkonium wave function. In a previous publication~\cite{Eichten:1995ch}, we tabulated the values at the origin of (the absolute square of) the radial wave function (for $s$-wave levels), or its first nonvanishing derivative (for orbitally excited levels), for a selection of quarkonium potentials that gave reasonable accounts of the $\jpsi$ and $\Upsilon$ spectra then known. There we examined the Cornell Coulomb-plus-linear potential~\cite{Eichten:1978tg, *Eichten:1979ms}, Martin's power-law potential~\cite{Martin:1980jx}, Richardson's QCD-inspired potential~\cite{Richardson:1978bt}, and a second QCD-inspired potential due to Buchm\"uller and Tye~\cite{Buchmuller:1980su}.

The QCD-inspired potentials incorporate running of the strong coupling constant \alphas\, using the perturbative-QCD evolution equation at leading order and beyond. But at distances relevant for confinement, perturbation theory ceases to be a reliable guide. It is now widely held, following Gribov~\cite{Gribov:1999ui}, that \alphas\ approaches a critical, or frozen, value at long distances (low energy scales), as a result of quantum screening. Recently we incorporated the spirit of this insight into a new version of the Coulomb-plus-linear form that we call the \emph{frozen-\alphas\ potential}~\cite{Eichten:2019gig}, which we employed in a prospectus for the $(c\bar{b})$ spectrum. The purpose of this note is to record calculations for the wave functions at the origin in all three of the quarkonium families, $\jpsi, B_c$, and $\Upsilon$.

For bound states in a central potential, 
the Schr\"{o}dinger wave function separates neatly into
radial and angular pieces, as $\Psi_{n\ell m}(\vec{r}) = R_{n\ell}(r)Y_{\ell m}(\theta,\phi)$,
where $n$ is the principal quantum number, $\ell$ and $m$ are the
orbital angular momentum and its projection, $R_{n\ell}(r)$ is the radial
wave function, and $Y_{\ell m}(\theta,\phi)$ is a spherical harmonic with normalization
 $\int d\Omega\,Y_{\ell
m}^*(\theta,\phi)Y_{\ell^\prime m^\prime}(\theta,\phi) =
\delta_{\ell \ell^\prime}\,\delta_{m m^\prime}$.    From the normalization condition $\int{ d^3\vec{r}\, |\Psi_{n\ell m}(\vec{r})|^2} = 1$,
it follows that $\int_0^\infty r^2 dr |R_{n\ell}(r)| = 1$.
The value of the radial wave function, or its first nonvanishing derivative, at the origin,
\begin{equation}
	R_{n\ell}^{(\ell)}(0)\equiv \left.\frac{d^{\,\ell}R_{n\ell}(r)}
	{dr^\ell}\right|_{r=0}\;\;\;,
	\label{wvfc}
\end{equation}
is required to evaluate meson decay constants and the parton-fragmentation contribution to production rates.

The long-range part of the frozen-\alphas\ potential has the standard Cornell linear form. To obtain the Coulomb piece, we converted the four-loop running of $\alphas(q)$ in momentum space~\cite{Chetyrkin:2004mf,*Czakon:2004bu} to the behavior in position space using the method of~\cite{Jezabek:1998wk}, and enforced saturation of $\alphas(r)$ at long distances, as detailed in Ref.~\cite{Eichten:2019gig}. Values of $\alphas(r)$ in a form convenient for interpolation are presented in the Appendix to that article. 

For each quarkonium level, we show in Table~\ref{tab:wfos} the computed centroid of the mass along with the wave function at the origin. For each family, we have included not only the narrow levels, but also states above flavor threshold that might be useful in assessing cascade contributions or the production of discrete excited levels. Comparing with Tables I, II, and III of Ref.~\cite{Eichten:1995ch}, we find that the new values lie comfortably within the range of our earlier estimates~\footnote{The strong Coulomb term of the original Cornell potential is reflected in spatially compact states, and large values of the wave functions at the origin.}. We recommend the new results presented here as reliable modern reference values.

\begin{acknowledgments}
This work was supported by Fermi Research Alliance, LLC under Contract No. DE-AC02-07CH11359 with the U.S. Department of Energy, Office of Science, Office of High Energy Physics.
\end{acknowledgments}

 \begin{table*}
 \caption{Absolute squares of radial wave functions or their first nonvanishing derivatives at the origin for quarkonium states. The masses are centroids of the indicated states, calculated in the frozen-\alphas\ potential with $m_c = 1.84\gev$ and $m_b = 5.19\gev$.\label{tab:wfos}}
 \begin{ruledtabular}
 \begin{tabular}{ccccccc}
 \multicolumn{1}{c}{Level} & \multicolumn{2}{c}{$c\bar{c}$} & \multicolumn{2}{c}{$c\bar{b}$}& \multicolumn{2}{c}{$b\bar{b}$}\\[-6pt]
  &  Mass [GeV] & $|R^{(\ell)}_{n\ell}|^2$ & Mass [GeV] & $|R^{(\ell)}_{n\ell}|^2$ & Mass [GeV] & $|R^{(\ell)}_{n\ell}|^2$\B \\ \hline
  $1S$ & $3.0687$ & $1.0952\gev^3$ & $6.3155$ & $1.9943\gev^3$ & $9.4425$ & $5.8588\gev^3$\T\\
  $2P$ & $3.5029$ & $0.1296\gev^5$ & $6.7517$ & $0.3083\gev^5$ & $9.8827$ & $1.6057\gev^5$\\[-2pt]
  $2S$ & $3.6790$ & $0.6966\gev^3$ & $6.8860$ & $1.1443\gev^3$ & $10.0159$ & $2.8974\gev^3$\\
  $3D$ & $3.7983$ & $0.0329\gev^7$ & $7.0179$ & $0.0986\gev^7$ & $10.1448$ & $0.8394\gev^7$\\[-2pt]
  $3P$ & $3.9554$ & $0.1767\gev^5$ & $7.1539$ & $0.3939\gev^5$ & $10.2550$ & $1.8240\gev^5$\\[-2pt]
  $3S$ & $4.1079$ & $0.5951\gev^3$ & $7.2732$ & $0.9440\gev^3$ & $10.3639$ & $2.2496\gev^3$\\
  $4F$ & $4.0419$ & $0.01317\gev^9$ & $7.2348$ & $0.0493\gev^9$ & $10.3454$ & $0.6643\gev^9$\\[-2pt]
  $4D$ & $4.1873$ & $0.06923\gev^7$ & $7.3630$ & $0.1989\gev^7$ & $10.4461$ & $1.5572\gev^7$\\[-2pt]
  $4P$ & $4.3297$ & $0.2106\gev^5$ & $7.4864$ & $0.4540\gev^5$ & $10.5451$ & $1.9804\gev^5$\\[-2pt]
  $4S$ & $4.4685$ & $0.5461\gev^3$ & $7.5963$ & $0.8504\gev^3$ & $10.6421$ & $1.9645\gev^3$\\
  $5G$ & $4.2573$ & $0.00750\gev^{11}$ & $\cdots$ & $\cdots$ & $10.5153$ & $0.7392\gev^{11}$\\[-2pt]
  $5F$ & $4.3937$ & $0.03740\gev^9$ & $\cdots$ & $\cdots$ & $10.6100$ & $1.7228\gev^9$\\[-2pt]
  $5D$ & $4.5284$ & $0.1074\gev^7$ & $\cdots$ & $\cdots$ & $10.7031$ & $2.2324\gev^7$\\[-2pt]
  $5P$ & $4.6607$ & $0.2389\gev^5$ & $\cdots$ & $\cdots$ & $10.7947$ & $2.1175\gev^5$\\[-2pt]
  $5S$ & $4.7900$ & $0.5160\gev^3$ & $\cdots$ & $\cdots$ & $10.8843$ & $1.7990\gev^3$\\
  $6H$ & $\cdots$ & $\cdots$ & $\cdots$ & $\cdots$ & $10.6668$ & $1.1071\gev^{13}$\\[-2pt]
  $6G$ & $\cdots$ & $\cdots$ & $\cdots$ & $\cdots$ & $10.7568$ & $2.4623\gev^{11}$\\[-2pt]
  $6F$ & $\cdots$ & $\cdots$ & $\cdots$ & $\cdots$ & $10.8457$ & $3.0936\gev^9$\\[-2pt]
  $6D$ & $\cdots$ & $\cdots$ & $\cdots$ & $\cdots$ & $10.9332$ & $2.8903\gev^7$\\[-2pt]
  $6P$ & $\cdots$ & $\cdots$ & $\cdots$ & $\cdots$ & $11.0194$ & $2.2430\gev^5$\\[-2pt]
  $6S$ & $\cdots$ & $\cdots$ & $\cdots$ & $\cdots$ & $11.1036$ & $1.6885\gev^3$\\
  $7I$ & $\cdots$ & $\cdots$ & $\cdots$ & $\cdots$ & $10.8058$ & $2.1639\gev^{15}$\\[-2pt]
  $7H$ & $\cdots$ & $\cdots$ & $\cdots$ & $\cdots$ & $10.8919$ & $4.50055\gev^{13}$\\[-2pt]
  $7G$ & $\cdots$ & $\cdots$ & $\cdots$ & $\cdots$ & $10.9771$ & $5.3196\gev^{11}$\\[-2pt]
  $7F$ & $\cdots$ & $\cdots$ & $\cdots$ & $\cdots$ & $11.0614$ & $4.7389\gev^9$\\[-2pt]
  $7D$ & $\cdots$ & $\cdots$ & $\cdots$ & $\cdots$ & $11.1446$ & $3.5411\gev^7$\\[-2pt]
  $7P$ & $\cdots$ & $\cdots$ & $\cdots$ & $\cdots$ & $11.2266$ & $2.3600\gev^5$\\[-2pt]
  $7S$ & $\cdots$ & $\cdots$ & $\cdots$ & $\cdots$ & $11.3066$ & $1.6080\gev^3$\B
 \end{tabular}
 \end{ruledtabular}
 \end{table*}




\bibliography{wfo19}

\end{document}